# Predicting Autonomous Vehicle Collision Injury Severity Levels for Ethical Decision Making and Path Planning

James E. Pickering, Aston University, Birmingham UK
Keith J. Burnham, University of Wolverhampton, UK

**ABSTRACT**

Developments in autonomous vehicles (AVs) are rapidly advancing and will in the next 20 years become a central part to our society. However, especially in the early stages of deployment, there is expected to be incidents involving AVs. In the event of AV incidents, decisions will need to be made that require ethical decisions, e.g., deciding between colliding into a group of pedestrians or a rigid barrier. For an AV to undertake such ethical decision making and path planning, simulation models of the situation will be required that are used in real-time on-board the AV. These models will enable path planning and ethical decision making to be undertaken based on predetermined collision injury severity levels. In this research, models are developed for the path planning and ethical decision making that predetermine knowledge regarding the possible collision injury severities, i.e., peak deformation of the AV colliding into the rigid barrier or the impact velocity of the AV colliding into a pedestrian. Based on such knowledge and using fuzzy logic, a novel nonlinear weighted utility cost function for the collision injury severity levels is developed. This allows the model-based predicted collision outcomes arising from AV peak deformation and AV-pedestrian impact velocity to be examined separately via weighted utility cost functions with a common structure. The general form of the weighted utility cost function exploits a fuzzy sets approach, thus allowing common utility costs from the two separate utility cost functions to be meaningfully compared. A decision-making algorithm, which makes use of a utilitarian ethical approach, ensures that the AV will always steer onto the path which represents the lowest injury severity level, hence utility cost to society. The developed system is known as the ethical steering control system (ESCS). It is further investigated how the number of pedestrians and occupants on-board the AV influence the ethical decision making. The results presented in this research highlight the effectiveness of the ESCS on-board future AVs, with the potential to minimise the severity of AV incidents and/or number of fatalities.

## 1.0 ETHICAL STEERING CONTROL SYSTEM (ESCS)

In this research, the ethical dilemma involved with autonomous vehicles (AVs) is investigated by analogy with the trolley problem (Lin, 2013). The scenario of interest is illustrated in Figure 1, in which an AV (containing two occupants) enters an unavoidable collision. The ethical dilemma involves the AV having to make a decision, i.e., to remain on course and collide into a rigid barrier or take steering action and collide into two pedestrians. Both decisions will involve the AV applying emergency brakes to reduce the severity of the collision. The following question is posed: Assuming the utility of all persons involved in the collision scenario to be equal, and knowing this initial information, what steering action should be taken? In (Bonnefon, Shariff and Rahwan, 2015), participants of a survey were asked their views about scenarios similar to that in Figure 1. However, in this survey, the collision outcomes were limited to a binary life-or-death matter. Only two collision outcomes were considered, i.e., life or death. The majority (75%) of the survey participants selected for the AV to steer into the path of least fatalities. Such an approach is known as a utilitarian ethical approach, hence minimising the utility cost to society. Alternatively, if the AV was to remain on course and take no action (i.e., applying no steering action), this is known as a deontological approach. Based on Figure 1, the question is then: if the collision output could be predetermined, would the path which leads to the lowest cost to society be selected? In this paper the above two ethical approaches are investigated and their efficacies are compared.

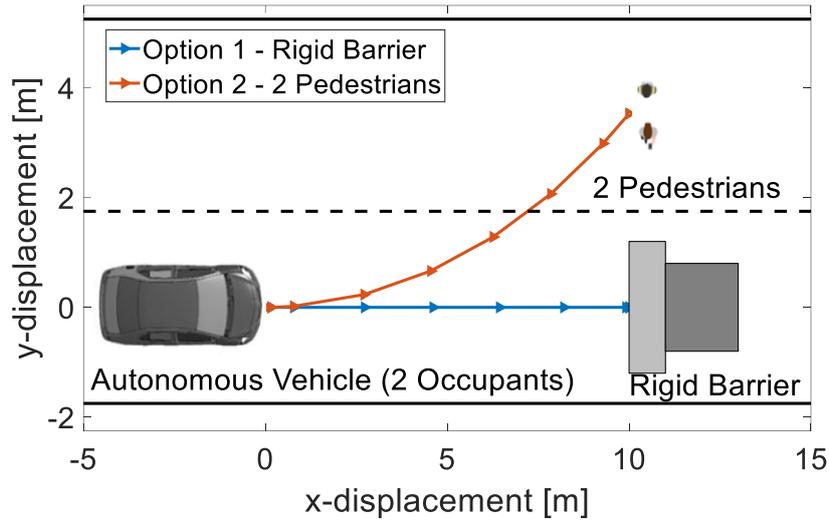

**Figure 1:** *Autonomous Vehicle containing Two Occupants Faced with an Ethical Dilemma: Option 1 to Steer into 2 Pedestrians or Option 2 to Remain On-Course and Collide with a Rigid Barrier*

Considering life and death outcomes for AV collision scenarios, which are analogous to the trolley problem, introduces an interesting ethical dilemma. This research aims to extend the considerations of such scenarios further by introducing collision injury severity levels, e.g., determining the collision injury severity of a collision with a pedestrian and an AV collision with a rigid barrier. If the collision injury severity levels can be predetermined for the two collision options given in Figure 1, the preferred utilitarian ethical approach would almost always be applied. Hence, a decision to steer the AV into the path of least injury severity level, or lowest fatalities, would be selected.

To determine the collision injury severity levels, a mathematical model of the dynamic motion of an AV motion is developed. This will allow the impact velocity of the collision events to be predicted, and how this contributes to the collision injury severity level. This allows the effect of the following properties on the collision 'target' upon applying the two ethical approaches to be explored:

  i. AV initial velocities: $12m/s$, $16m/s$ and $20m/s$,
  ii. Number of occupants occupying the front seats of the AV: zero with an unladen mass of $1247kg$, one with a laden mass of $1327kg$ and two with a laden mass of $1407kg$
  iii. Number of pedestrians: 0 to 4

Therefore, based on knowing the above information, along with the mathematical model of the AV motion, further research is undertaken to allow the injury severities of the collision outcomes to be determined. For example, given the scenario in Figure 1, how will the above properties affect the peak deformation of the AV and the level of occupant injury severity? Similarly, how will the AV to pedestrian impact velocity affect the injury severity level? The intention is for such a decision- making approach to be used on-board an actual AV in real-time, such that the collision injury severities can be predetermined. The collision injury severities can then be used to make ethical decisions, e.g., applying a utilitarian ethical approach to steer into the path of least severity. The name of the control system developed here is known as the ethical steering control system (ESCS).

## 1.1 Literature Review

Some initial research on this topic has already been undertaken by the authors, see e.g., (Pickering, 2020). The initial 'spark' of interest for this research came from a paper published in 2016 by (Goodall, 2016). In that paper, a collision scenario is introduced whereby there are multiple possible collision outcomes. The question is introduced – which path should the AV take in selecting the collision target? Prompted by the paper by Goodall (2016), work by the authors started on developing a model-to-decision (M2D) approach as a possible solution to the problem. This M2D approach involves developing mathematical models of the collision scenarios to allow the injury severity levels of the potential collision options to be determined. Based on predetermining the collision injury levels, ethical decision-making algorithms are applied, e.g., use of the utilitarian ethical approach to steer into the path of least injury severity. In (Pickering, et al., 2018), a mathematical model was developed for a highway collision scenario, i.e., exploring AV rear-end collisions. Using multi-criteria decision-making methods, an ethical decision-making algorithm was developed using an analytical hierarchy process (AHP). It was determined that via the use of mathematical models, the collision path of least severity could be determined. This work was extended in (Gilbert et al., 2021), where further work was undertaken, e.g., using a steering model and a sensitivity study. In the research, it was found that various parameters (e.g., AV mass) of the AV ahead and behind in a highway lane scenario can change the lane selection. Regarding the scenario to be investigated in this research (i.e., AV collision into pedestrian(s) or a rigid barrier), an initial study was undertaken in (Pickering, Podsiadly and Burnham, 2019). In that paper, a mathematical framework was proposed, which enabled the prediction of injury severity levels via a model-based approach. As in (Gilbert et al., 2021), it was demonstrated that the collision path of least injury severity can be selected.

In (Gerdes and Thornton, 2015) and (Thornton, Pan and Gerdes, 2016), the authors explored the ethical theories of deontology and consequentialism to provide guidance for responsible programming of an AV. Although the research does not involve ethical decision making in the event of a collision, the ethical approach is of interest. A model predictive control (MPC) approach was used which minimised the consequential costs subject to deontological constraints. Within the approach, path planning, vehicle occupant comfort and traffic laws are priorities of the cost function within the MPC, whilst obstacle avoidance and vehicle slew rate limits enter as constraints. In (Evans, et al., 2020), a computational approach has been developed that accommodates moral AV collision scenarios. This considered the morality demands and what road users may expect, with the research offering an evaluation tool for the social acceptability of an AV ethical decision-maker.

## 2.0 STEERING CONTROL SYSTEM

This section details the three stages which are to be developed as part of the ESCS for an AV:

i. Kinematic bicycle model
ii. Longitudinal dynamics model
iii. Velocity control

MATLAB and Simulink 2022a have been used throughout for modelling and simulation. The approach developed here is similar to that developed in (D'Souza, Burnham and Pickering, 2022), however, in that research the highway three-lane scenario was considered.

### 2.1 Modelling Assumptions

With reference to the scenario in Figure 1, Section 1.0, the following assumptions are considered in the development of the approach:

- Each of the road lanes have a 3.5metre width and only a 'straight' road is considered
- Detailed aerodynamic and friction forces are not considered in the modelling
- The AV considered in the simulation is initially located in the centre of the respected lane
- The steering angle remains constant throughout the course of the simulated steering action
- Steering angles beyond $\pm 10°$ are not considered
- AVs are of Level 5 autonomy, i.e., fully autonomous.
- All occupants and pedestrians are of the same utility, i.e., equally valued to society

### 2.2 Kinematic Bicycle Model

A kinematic bicycle model is used in this research to represent the AV motion in a 2D model, i.e., the $x$- and $y$-axis. Based on (Chandran and Mathew, 2007), the following equations of motion describe the kinematic bicycle model used in this research:

$$\dot{x} = v \cos(\theta) \tag{1}$$

$$\dot{y} = v\sin(\theta) \tag{2}$$

$$\dot{\theta} = \frac{v}{L}\tan(\gamma) \tag{3}$$

where $L$ denotes the distance between the front and rear wheel axis of the AV, $\gamma$ denotes the AV steering angle output, $\theta$ denotes the AV heading/yaw angle, $\dot{x}$ is the velocity in the $x$-axis, $\dot{y}$ is the velocity in the $y$-axis and $\dot{\theta}$ is the angular velocity of the AV. Based on Equations (1) to (3), the AV has two inputs (i.e., vehicle speed, $v$ and steering angle, $\gamma$) and three outputs (i.e., $x$-axis velocity $\dot{x}$, $y$-axis velocity $\dot{y}$ and angular velocity $\dot{\theta}$ of the AV). This is captured in the Simulink diagram given in Figure 2 (right-hand-side). Integrators are used on the three outputs so that the $x$-axis displacement, $y$-axis displacement and heading/yaw angle $\theta$ outputs of the AVs motion can be captured, as these are of interest in the modelling and simulation.

## 2.3 Longitudinal Dynamic Model

In this Section, the velocity control element of the steering control system is developed. The aim of the velocity control is to minimise this value upon a collision being detected. This is important as upon detecting a collision, the AV will immediately apply the emergency brakes to minimise the collision velocity and thus, the collision severity.

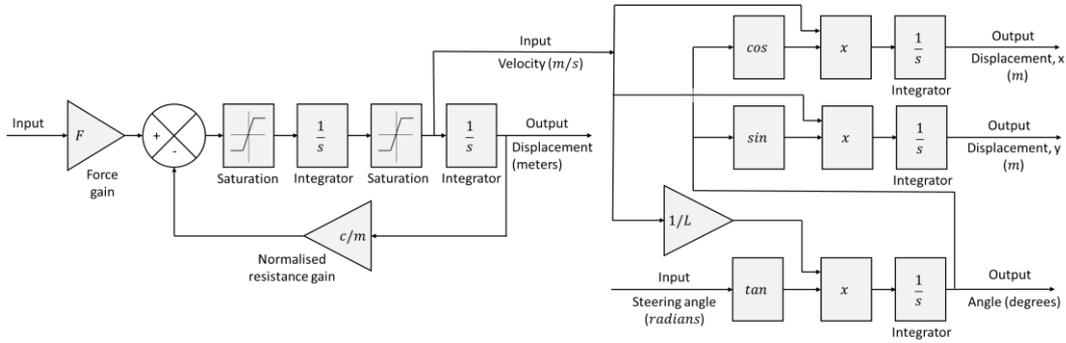

**Figure 2:** *Dynamic Model for the AV Velocity Control (Left-Hand-Side) and Kinematic Bicycle Model (Right-Hand-Side)*

To model the deceleration properties of the AV (i.e., when the brakes are applied), a dynamic model for the AV longitudinal motion is developed (assuming ideal actuators). This is used to capture the deceleration properties of the AV upon applying the brakes, and the corresponding collision impact velocity. The following transfer function is used to capture the dynamics of the AV longitudinal model:

$$G(s) = \frac{X(s)}{F(s)} = \frac{1}{ms^2 + cs} = \frac{1}{s(sm + c)} \tag{4}$$

where $m$ denotes the mass of the AV, $c$ denotes the combined rolling resistance and aerodynamic drag. The AV longitudinal motion given by Equation (4) is captured in the schematic given in Figure 2 (left-hand-side). The maximum velocity of the AV is introduced to the left saturation block in Figure 2. The maximum acceleration and the maximum deceleration are introduced to the right saturation block in Figure 2. The input in Figure 2 is set-up such that a value of 1 implies maximum acceleration and a value of -1 implies maximum deceleration of the AV. The input multiplies the force, denoted $F$, with this 'driving' the AV longitudinal motion. An initial velocity of the longitudinal motion model is introduced to the left-hand integrator.

## 2.4 Velocity Control

To control the velocity of the AV's motion, a closed-loop control system is utilised, see Figure 3. For this, a simple proportional control gain, denoted $P$, has been used. As the output from the proportional control gain is effectively force (as an ideal actuator is assumed), a force saturation block is used to limit the maximum force supplied. Recall from Section 2.3 that the initial condition (velocity) of the vehicle motion is introduced to the left integrator. Upon a collision scenario being detected, the reference of the AV velocity control is set to a value of zero. Reducing the collision impact velocity is zero is highly desirable, due to the squared velocity term in the kinematic energy formula, i.e., $E = 0.5mv^2$. Of course, reducing the impact velocity to zero is desired, however, in many scenarios this will not be possible, and a collision will occur. The simulation is stopped once the collision event occurs, i.e., the AV colliding into a barrier or a group of pedestrians.

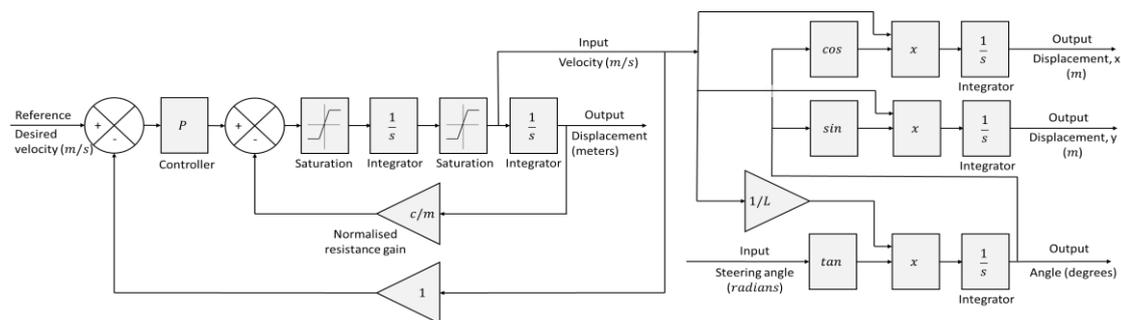

**Figure 3:** *Velocity Control of Longitudinal Bicycle Model*

## 2.5 Illustrative Example

The developed steering control system is now simulated using the initial scenario that is detailed in Figure 1, Section 1.0. In Section 3.2, the properties of a Toyota Yaris Sedan (TYS) 2010 is used to develop the collision injury severity levels, therefore, the simulation in this illustrative example will be based on actual vehicle properties. The velocity control of a longitudinal bicycle model given in Figure 3 is now simulated with the following properties:

- Kinematic bicycle model: distance between the front and rear wheel axis, $L$, of 2.55 metres and two steering angle outputs, $\gamma$, of 0 and 0.15 radians (0 and 8.59 degrees)
- Longitudinal motion model: the laden mass of the AV is $1407kg$ (baseline of the vehicle is 1247kg, and two occupants are added with a mass of $80kg$ each), combined rolling resistance and aerodynamic drag is $140Ns/m$, the maximum velocity of the AV is $107mph$ ($47.8m/s$), maximum acceleration is $8.5m/s^2$, the maximum deceleration is $5m/s^2$, the force input is $10600N$ and an initial velocity of $20m/s$ ($44.7mph$) has been set as the initial condition for the vehicle motion model
- Velocity control: proportional control gain, denoted $P$, given a value of 70

The use of the simulation tool will be initially demonstrated using the values given above. The graphical outputs from the simulation are given in Figure 4, where it is clear from the lower plot that the initial velocity is $20m/s$. After applying the brakes (i.e., the reference is now $0\ m/s$), the vehicle decelerates over the distance to the collision targets 10 meters away. Using the simulation, it is possible to determine the velocity upon impact between the AV and the collision target (pedestrians or rigid barrier). As the $x$-displacement between the AV and target is approximately 10 metres, the impact velocity can be determined using the longitudinal motion model outputs, i.e., in this case, $17.32m/s$. Based on the scenario (i.e., given in Figure 1, Section 1.0) and the determined pedestrian impact velocity and peak deformation of the AV into the rigid wall, the following question arises; which of the two collision targets would be the 'best' option? In Sections 3.0 and 4.0, the collision injury severity of such outcomes will be covered, with this being developed as part of the ethical decision maker. Based, on this, an estimate of the collision injury severity of each collision outcome will be determined, and this then will be used in the choice of 'best' steering action.

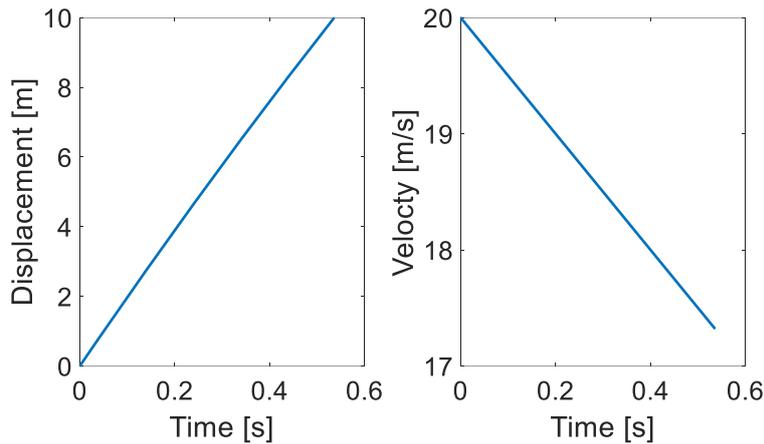

*Figure 4: Time Versus Displacement and Time Versus Velocity upon the Steering Control System being Deployed, i.e., Aiming to Reduce Velocity to Zero and Taking Steering Action*

## 3.0 COLLISION INJURY SEVERITY

In the previous section, a steering control system was developed, with this effectively providing information in terms of the AV collision mass and velocity. Such properties will influence the collision injury severity and fatality risk. Hence, in this section, the aim is to determine the collision injury severity for a pedestrian concerning impact velocity and also the peak deformation of an AV. For the evaluation of the pedestrian impact velocity, historic data is used. The AV injury severity will be based on the value of the peak deformation. The designed peak deformation will be considered, i.e., the deformation of the crumple zones without intrusion into the passenger cell (in this research, a value of $0.5900m$ for the vehicle is used). Peak deformation values below this will be considered low risk and values above this will be considered high risk (and increasing with peak deformation). To understand the effect of AV mass and velocity on peak deformation, a mathematical model is developed. The mathematical model is based on captured finite element (FE) data from a Toyota Yaris Sedan 2010 vehicle. The FE model is of high fidelity, but the model developed here will be of low fidelity, i.e., a mass-spring second linear order ordinary differential equation model. A least squares approach is used to capture the stiffness coefficient of the spring component, i.e., applying this to the deformation versus force data. Due to the low fidelity nature of the model, this can be used in real time to predetermine the peak deformation prior to a collision (based on estimating/predicting the AV laden mass and collision velocity).

### 3.1 Pedestrian Impact Velocity

In this Section, the pedestrian injury severity and fatality risk is now detailed for a single AV colliding into a pedestrian. Based on the literature review, the following three areas influence the injury severity and probability of survival:

- Age of the pedestrian
- Vehicle size/type
- Collision impact velocity

To simplify the modelling and ethical considerations the age of pedestrians and vehicle size/type are considered to be fixed in this research. Instead, the focus here will be on collision impact velocity only, and the effect of this on the injury severity and fatality risk. Of interest, in (Richards, et al, 2010), human driven vehicles travelling at higher velocities are more likely to be involved in serious pedestrian accidents than those travelling at lower velocities. In (Pitt et al, 1990), the research findings state that the collision impact velocity has a significant effect on the force of the impact, with this being the most direct variable to determine injury severity. There are various studies involving the vehicle-pedestrian impact velocity, with various findings from the author's, see (Tefft, 2013), (Roudsari et al., 2004), (Fredriksson, Rosén and Kullgren, 2010), (Davis, 2001) and (Rosén and Sander, 2009).

In (Tefft, 2013), a study in the USA was undertaken, where it was determined that 31.7% of severe injuries occurred at vehicle collision impact velocities below 32 $km/h$ (i.e., 8.8889$m/s$ or 19.8839$mph$). A similar project was undertaken in the UK, see (Ashton, 1980), where it was determined that 32.7% of severe vehicle collision injuries occur at velocities below 30$km/h$ (i.e., 8.3333 $m/s$ or 18.6411$mph$). Although both studies were published at different times (2013 and 1980, respectively), the data suggests that for severe injuries to be avoided, pedestrian impact velocities of 30$km/h$ may not be low enough. However, it is to be expected that since the study published in 1980 (Ashton, 1980), vehicle design has improved, therefore, impact velocities of 30 $km/h$ are currently likely to be too high. There are also expected to be significant differences between USA and UK vehicle design, e.g., vehicle size and type of safety systems. In (Cuerden, Richards and Hill, 2007), a UK project known as 'On the Spot' (OTS) collected accident data from 500 road accidents per year starting in 2000 and worked closely with both the Nottinghamshire and Thames Valley Police Forces. The purpose of the research was to examine the case of pedestrian injury severity, where the authors used the UK Governments accepted descriptions of slight, severe and fatal, see (Department of Transport, 2017). Of the incidents recorded, 108 involved pedestrians and vehicles. The results from the study are summarised in Table 1, with the percentage of injuries arising over a range of impact velocities. Out of the 108 incidents and considering the pedestrians' head, these accounted for 70% of injuries and 90% of fatalities, involving the vehicles' A-pillar where the collision impact took place. In terms of the slight injuries, most of these were to the lower leg and knee due to the pedestrian impact with the vehicles front bumper. Some 85% of the incidents with pedestrians occurred at impact velocities lower than 50$km/h$ (i.e., 13.8889 $m/s$ or 31.0686 $mph$), where there is a lower probability of fatality. Whilst the captured data in Table 1 is small (108 incidents), the findings are in agreement with the earlier study carried out by Ashton and MacKay, see (Ashton and MacKay, 1979), where in Table 1 it is determined that 33% of severe injuries occur below 30$km/h$ (i.e., 8.3333 $m/s$ or 18.6411$mph$).

Figure 5 illustrates the findings from (Kröyer, 2015) and (Rosén and Sander, 2009), where the graphical outputs of mean travelling velocity and the collision impact velocity involving a 40-year-old pedestrian versus percentage possibility of fatality are plotted. There are differences between the two cures, as the driver typically brakes to reduce the collision impact velocity. Of interest, the curve in Figure 5 closely matches the data presented in Table 1. For example, in Table 1, an impact velocity of 50 – 60$km/h$ (i.e., 31.0686$mph$ - 37.2623$mph$) gives a fatal injury percentage of 11% which closely matches the 60$km/h$ value given in Figure 5.

**Table 1:** *Highlighting Severity of Pedestrian Injury and Corresponding Impact Velocity (Cuerden, Richards and Hill, 2007)*

| Impact Velocity [km/h] | Slight Injury [%] | Severe Injury [%] | Fatal Injury [%] |
|---|---|---|---|
| 0 – 10 (6.2137$mph$) | 100 | - | - |
| 10 (6.2137 $mph$) – 20 (12.4274 $mph$) | 60 | 40 | - |
| 20 (12.4274 $mph$) – 30 (18.6411 $mph$) | 67 | 33 | - |
| 30 (18.6411 $mph$) – 40 (24.8548 $mph$) | 53 | 43 | 4 |
| 40 (24.8548 $mph$) – 50 (31.0686 $mph$) | 22 | 65 | 13 |
| 50 (31.0686 $mph$) – 60 (37.2623 $mph$) | - | 89 | 11 |
| 60+ (37.2623 $mph$) | - | 50 | 50 |

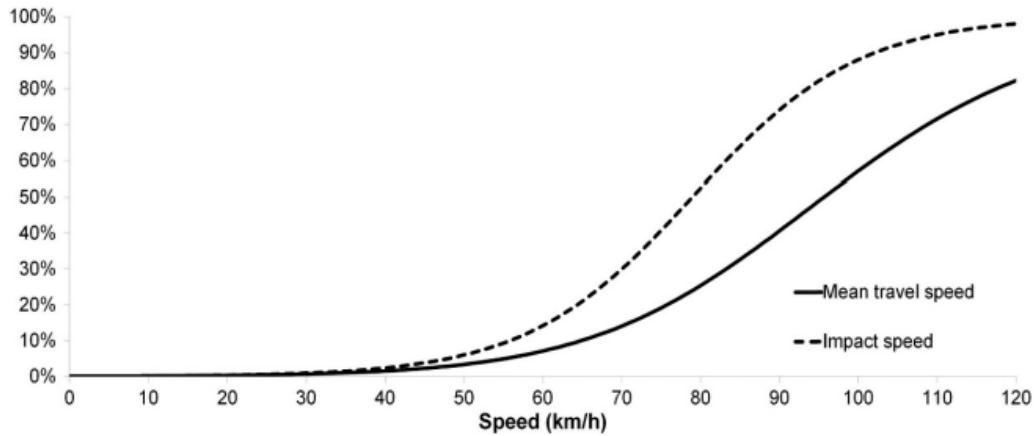

**Figure 5:** *Fatality Risk Curve Based on Impact Velocities, see (Rosén and Sander, 2009) and Fatality Risk Curve Based on Mean Travelling Velocity for a 40-year-old victim, see (Kröyer, 2015)*

### 3.2 Vehicle Peak Deformation

In this Section, the details of a high-fidelity FE collision simulation of a single vehicle into a rigid barrier are initially given, see Figure 6. The input to the FE collision simulation is derived from the initial condition due to the collision impact velocity and vehicle laden mass and the outputs are deformation versus time, acceleration versus time and collision deformation energy. Note that displacement is interpreted as deformation of the crumple zone is this research. The FE model was developed through the process of reverse engineering at the National Crash Analysis Centre (NCAC), George Washington University (GWU), see (Marzougui et al., 2012). The FE model is based on the 2010 Toyota Yaris Sedan (TYS), as initially illustrated in Figure 1. The 2010 TYS FE model is available for research as an open-access source from the Centre for Collision Safety and Analysis (CCSA) website, see (Centre for Collision Safety and Analysis, 2017).

The FE collision model is used as a surrogate for an actual vehicle collision. In the absence of an actual vehicle collision, the FE model goes some way to explain the physical phenomena, i.e., the conservation of momentum and energy, the coefficient of restitution, the collision forces and the structural stiffness. It should be noted that the purpose of this section is not to develop the FE model, but rather to gain an understanding of the cause-and-effect properties of a full-frontal collision, i.e., input (cause) and output (effect) properties. The computation and analysis of the FE model were undertaken using the proprietary LS-DYNA explicit computer solver. The 2010 TYS FE model will be used as the nominal benchmark vehicle collision model in this work. The FE collision model is then effectively taken as a surrogate for the actual vehicle collision system as this simulation provides the closest match to an actual realistic collision, see Table 2.

As the TYS FE model properties given in Table 2 are fixed (i.e., vehicle mass and velocity), there is a need to develop a low fidelity model that can be used in real time. The model should be able to capture the physical phenomena of the collision and allow for the physical properties of the vehicle to be investigated (i.e., vehicle mass and velocity). Using the mathematic model, the outputs of a given collision scenario can be predetermined, e.g., the peak deformation of the vehicle.

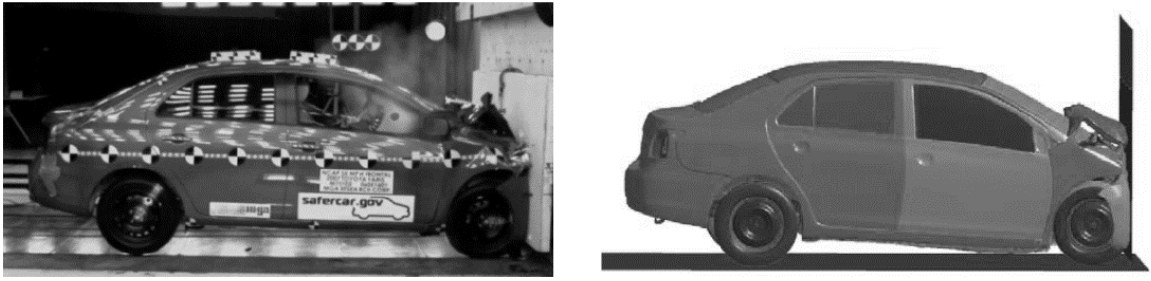

**Figure 6:** *Comparison of 2010 Toyota Yaris Front Collision Structure Deformation to the Finite Element Analysis Vehicle Model, see (Marzougui et al., 2012)*

**Table 2:** *Frontal Collision Output Results for Actual Data and Finite Element (Marzougui et al., 2012)*

|  | **Actual Vehicle** | **FE Model** |
| --- | --- | --- |
| Mass [kg] | 1271 | 1247 |
| Impact Velocity [m/s] | 15.6464 | 15.6464 |
| Post-Test Deformation [m] | 0.5170 | 0.5201 |
| Peak Deformation [m] | 0.5620 | 0.5625 |
| Designed Deformation Length [m] | --- | 0.5900 |
| Peak Acceleration [g] | ~ 52.00 | 55.04 |
| Collision Energy – Main Data [kJ] | --- | 149.1 |
| Collision Energy – Rebound Data [kJ] | --- | 7.560 |
| Collision Energy [kJ] | --- | 141.5 |
| Collision Duration [s] | --- | 0.0522 |

The modelling work to be undertaken makes use of a lumped parameter model (LPM) consisting of stiffness and mass values (i.e., a second order linear ordinary differential equation). For convenience and consistency, modelling of the LPM damping is not considered. This is justified as only the first quarter cycle of the dynamic response is considered. Considering the duration up to the peak deformation (which corresponds to the point where the velocity becomes zero) of the TYS, the time duration to reach this peak provides information of the force from which the collision energy may be derived (related to the area under the first quarter cycle). Therefore, the low fidelity LPM aims to capture the peak deformation of the FE TYS sufficiently accurately, whilst bearing in mind that all mathematical/numerical models are approximations.

The following 2nd order differential equation is used as a representation of a single Vehicle, denoted Vehicle $a$, colliding into a rigid barrier:

$$m_a \frac{d^2 x_a(t)}{dt^2} + k_a x_a(t) = f_a(t) \tag{5}$$

where $m_a$ denotes the vehicle laden mass, $k_a$ denotes the average structural stiffness value, $f_a$ denotes the applied force to the rigid barrier and $x_a$ denotes the structural deformation of the vehicle crumple zone. The coefficient for the spring stiffness constant $k_a$ in Equation (5) needs to be tuned, such that the simulation of the LPM adequately replicates the peak deformation of the FE simulation. A structural stiffness $k_a$ value for Equation (5) will be determined using a linear least squares regression model. This will make use of the baseline TYS FE data captured based on the US NCAP test, i.e., a nominal mass of $1247 kg$ and a velocity of $35 mph$ ($15.6464 m/s$). A linear regression single-section (straight-line) model of the force versus deformation plot, corresponding to a vehicle, denoted Vehicle $a$, is given by the equation of the following affine/linear form:

$$f_a = f_{p_a} + k_a \delta_a \tag{6}$$

where $f_{p_a}$ denotes, for Vehicle $a$, the crumple zone failure point (or buckling point) and $\delta_a$ denotes the progressive deformation of the vehicle's frontal longitudinal crumple zone. To estimate the failure point and the average stiffness value, a linear least squares (LLS) approach is applied to the force versus deformation data, which effectively determines the 'best' fit to the observed data in the sense of minimising the sum of squares of the 'vertical' deviations from each point from a straight-line segment, i.e., if a data point lies on the fitted line, then the vertical deviation is zero. The LLS general form is given by (Hsia, 1977):

$$\hat{\theta}_a = [\Phi_a^T \Phi_a]^{-1} \Phi_a^T x_a \qquad (7)$$

where $\hat{\theta}_a$ is the estimated parameter vector, hence $\hat{\theta}_a = \begin{bmatrix} f_{p_a} \\ k_a \end{bmatrix}$ and $\Phi_a$ is the observation matrix consisting of simulated values of the inputs (force, due to the initial collision velocity) and the simulated values of the outputs (deformation denoted $x_a$) which are the deformation values extracted from the TYS FE simulation data.

In Figure 7, the result of applying the LLS approach is given, where the single straight-line illustrates the fit to the TYS FE model force versus deformation data. The determined values for the failure point $f_{p_a}$ and structural stiffness $k_a$ can be found in Table 3. As a measure of the effectiveness of the estimated model obtained via application of the LLS approach to the FE collision data, integration is used to find the area under the force versus deformation plot (collision energy). Recall that the FE data is considered as a surrogate model so the use of the term 'error' is really interpreted here as a 'deviation'.

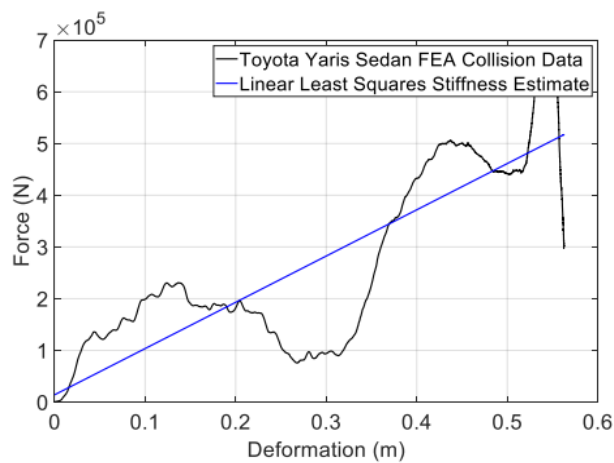

**Figure 7:** *Single Straight-Line Approximation using Linear Least Squares*

**Table 3:** *Estimate of Stiffness, Estimate of Failure Point and Area under the Graph*

| Estimate of stiffness $k_a$ [kN/m] | Estimate of failure point $f_{p_a}$ [kN] | Area under the graph [kJ] |
|---|---|---|
| 894.3 | 1.410 | 149.2 |

Equation (5) has been simulated using the values in Table 2 (i.e., impact velocity and vehicle laden mass) and Table 3 (i.e., estimate of stiffness). The results from the simulation are given in Figure 8, where the TYS FE collision data is compared to the single LPM. The key information from Figure 8 is captured in Table 4, i.e., peak deformation, collision energy and collision duration. The last column in Table 4 gives the magnitude of the discrepancy expressed as a per unit value between the FE simulation output and the simulated LPM output obtained from:

$$Discrepancy = \left[\frac{FE\ Data - LPM\ Data}{FE\ Data}\right] \quad (8)$$

Expressed as a magnitude, the LPM peak deformation matches closely to that of the FE data, with a discrepancy of 0.0386. The discrepancy in the collision energy value for the single LPM has a value of 0.0791. Based on the conservation of energy, the LPM with the failure point given in Table 3 produced a value above that expected. Therefore, the failure point was removed, with the comparison of the results given in Table 4. Without the failure point, a result is now produced which is more adequate and acceptable, i.e., a value of $152.6 kJ$ and a small discrepancy value of 0.0235.

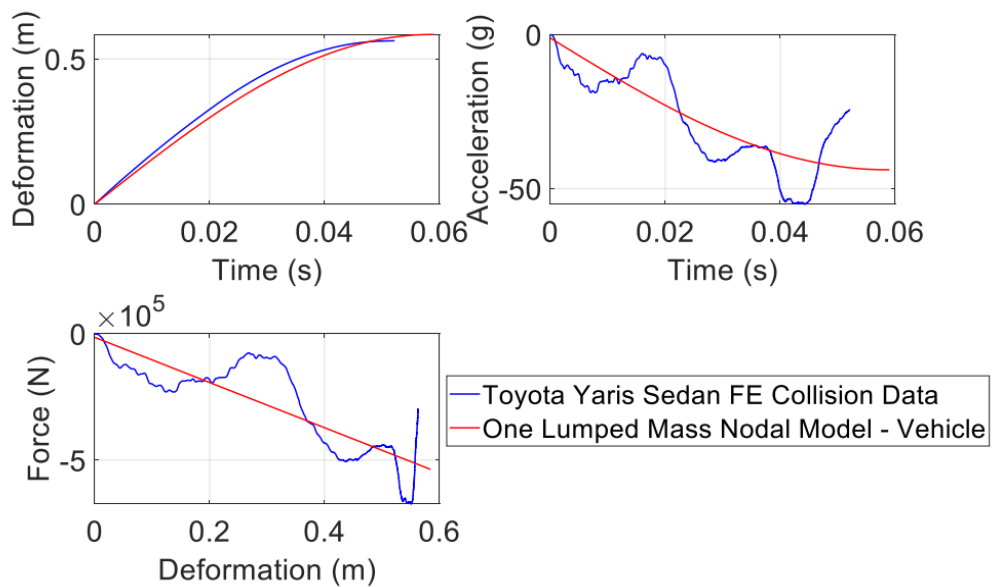

**Figure 8:** *Comparison of the Structural Properties of the FE Model to the One Lumped Parameter Model for Deformation Versus Time (top-left hand sub-plot), Acceleration versus Time (top-right hand sub-plot) and Force versus Deformation (bottom-left hand sub-plot)*

**Table 4:** *Comparison of Structural and Occupant Properties of the FE Model to the One Lumped Parameter Model*

|  | FE Model | Lumped Mass Model | Discrepancy [±] |
| --- | --- | --- | --- |
| Peak Deformation [$m$] | 0.5625 | 0.5842 | 0.0386 |
| Collision Energy, with Failure Point [$kJ$] | 149.1 | 160.9 | 0.0791 |
| Collision Energy, without failure point [$kJ$] | 149.1 | 152.6 | 0.0235 |
| Collision Duration [$s$] | 0.0522 | 0.0590 | 0.1303 |

## 4.0 ETHICAL DECISION MAKER

This Section will now detail the ethical decision maker. This incorporates data from the collision literature/modelling for a pedestrian and an occupant within a vehicle, see Sections 3.1 and 3.2. This data is used here to develop injury severity levels for the scenario given in Figure 1, Section 1.0., i.e., the AV colliding into a rigid barrier containing 0 to 2 occupants or a group of up 1 to 4 pedestrians. In this Section, the data will be mapped onto fuzzy sets, where a collision output (i.e., pedestrian impact velocity or AV occupant peak deformation) will have a degree of membership to two fuzzy sets. As there are two different collision outcomes, a common weighted utility cost function for collision injury severity is proposed. The weighted utility cost function is designed to be nonlinear in an attempt to avoid, if possible, the collision outcomes with high injury severity. Based on a general weighted utility cost function, see Sub-section 4.3, weighted utility cost functions with a common structure are developed for both collision outcomes (i.e., relating to AV peak deformation and pedestrian impact velocity). The common utility costs, or numerical outputs, from the two weighted utility cost functions are then available for use within ethical decision-making algorithms, i.e., the utilitarian and deontological approaches, which are further investigated. Illustrative examples for a range of scenarios are then presented, see Section 5, to demonstrate the operation of the developed ethical decision maker and the overall ESCS.

## 4.1 Fuzzy Sets for Vehicle Peak Deformation and Pedestrian Impact Velocity

For a single AV, denoted Vehicle $a$, involving a rigid barrier or a pedestrian(s), there are two features which correspond to the injury severity levels. These are peak deformation $\delta_a$ and pedestrian impact velocity $p_{v_a}$. The two injury severity levels, denoted $\delta_a$ and $p_{v_a}$, lie within the upper and lower ranges respectively:

$$s_{\delta_{1.00}} < s_{\delta_a} \leq s_{\delta_{5.00}} \tag{9}$$

$$s_{p_{v_{1.00}}} < s_{p_{v_a}} \leq s_{p_{v_{5.00}}} \tag{10}$$

in which $s_{\delta_{1.00}}$ and $s_{p_{v_{1.00}}}$ denote the lower bounds and $s_{\delta_{5.00}}$ and $s_{p_{v_{5.00}}}$ denote the upper bounds, respectively. The injury severity levels exist on fuzzy sets denoted $A$, $B$, $C$, $D$ and $E$, where fuzzy set $A$ corresponds to the highest injury severity and set $E$ corresponds to the lowest injury severity. The ranges given in Equations (9) to (10) are spanned by equally spaced crisp values between the minimum and maximum values for the 5 sets, see Table 5.

The fuzzy set for peak deformation ranges from $0.2681m$ to $0.8874\ m$ and utilises five fuzzy sets, denoted $E_\delta$ to $A_\delta$, respectively, with each fuzzy set $D_\delta$, $C_\delta$ and $B_\delta$ being identically triangular in shape, such that the peak for each fuzzy set is equally spaced and of magnitude unity. The peak deformation corresponding to the fuzzy sets is defined within the brackets for each case, as in Table 5. Note that $A_\delta(\delta_a)$ is $A_\delta(0.8874)$ and $E_\delta(\delta_a)$ is $E_\delta(0.2681)$, so that $C_\delta(\delta_a)$ is given by $(A_\delta(\delta_a) - E_\delta(\delta_a))/2$, similarly the intermediate cases for $B_\delta(\delta_a)$ and $D_\delta(\delta_a)$ are given by $(A_\delta(\delta_a) - C_\delta(\delta_a))/2$ and $(C_\delta(\delta_a) - E_\delta(\delta_a))/2$, respectively.

The fuzzy set for pedestrian impact velocity ranges from $6.7056m/s$ to $24.5872\ m/s$ and utilises five fuzzy sets, denoted $E_{p_v}$ to $A_{p_v}$, respectively, again with each intermediate fuzzy set being identically triangular in shape, such that the peak for each fuzzy set is equally spaced and of magnitude unity. The pedestrian impact velocity corresponding to the fuzzy sets is defined within the brackets for each case, as in Table 5. Note that $A_{p_v}(p_{v_a})$ is $A_{p_v}(24.5872)$ and $E_{p_v}(p_{v_a})$ is $E_{p_v}(6.7056)$, so that $C_{p_v}(p_{v_a})$ is given by $(A_{p_v}(p_{v_a}) - E_{p_v}(p_{v_a}))/2$, similarly the intermediate cases for $B_{p_v}(p_{v_a})$ and $D_{p_v}(p_{v_a})$ are given by $(A_{p_v}(p_{v_a}) - C_{p_v}(p_{v_a}))/2$ and $(C_{p_v}(p_{v_a}) - E_{p_v}(p_{v_a}))/2$, respectively.

**Table 5:** *Fuzzy Sets of Peak Deformation and Pedestrian Impact Velocity for Injury Severity*

| Peak Deformation [$m$] | | Pedestrian Impact Velocity [$m/s$] | |
|---|---|---|---|
| $s_{\delta_{5.00}}$ | $A_\delta(0.8874)$ | $s_{p_{v_{5.00}}}$ | $A_{p_v}(24.5872)$ |
| $s_{\delta_{4.00}}$ | $B_\delta(\delta_a)$ | $s_{p_{v_{4.00}}}$ | $B_{p_v}(p_{v_a})$ |
| $s_{\delta_{3.00}}$ | $C_\delta(\delta_a)$ | $s_{p_{v_{3.00}}}$ | $C_{p_v}(p_{v_a})$ |
| $s_{\delta_{2.00}}$ | $D_\delta(\delta_a)$ | $s_{p_{v_{2.00}}}$ | $D_{p_v}(p_{v_a})$ |
| $s_{\delta_{1.00}}$ | $E_\delta(0.2681)$ | $s_{p_{v_{1.00}}}$ | $E_{p_v}(6.7056)$ |

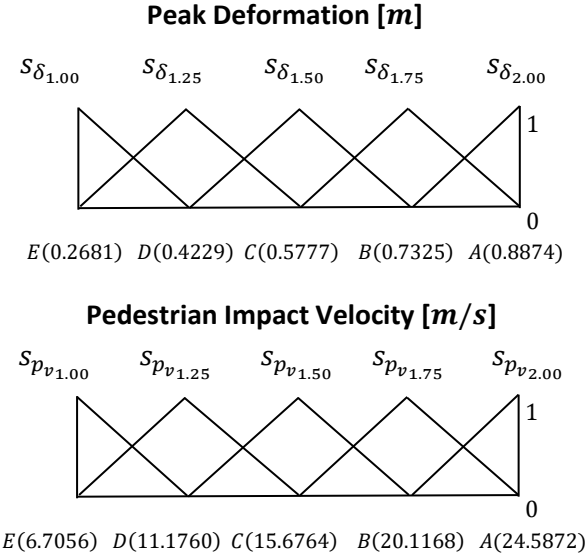

**Figure 9:** *Universe of Discourse for the Fuzzy Sets Corresponding to the Injury Severity for Peak Deformation $\delta_a$ and Pedestrian Impact Velocity $p_{v_a}$*

It is noted that less than five fuzzy sets could have been used. However, when considering the levels of pedestrian injury severity, with reference to the literature, see Section 3.1, it is convenient to use five fuzzy sets. In fact, it is the view of the authors that the proposed fuzzy sets ($A$, $B$, $C$, $D$ and $E$) for the impact velocity align well to the vehicle injury serveries. For example, when considering the vehicle, the fuzzy set '$C$' for peak deformation, peak head acceleration and peak chest acceleration are considered not fatal but would result in injuries. Considering the fuzzy set '$B$', the crisp values for the occupants of the vehicle and the pedestrians are considered to be venturing towards potentially fatal.

### 4.2 Membership Functions for Vehicle Peak Deformation and Occupant Impact Velocity

As in Section 4.1 and as discussed above, triangular fuzzy sets are used for the two features of peak deformation and pedestrian impact velocity, see Figure 9. The subscript notation $f$ is used as a general subscript to represent the two feature indices, i.e., for, $\delta_a$ and $p_{v_a}$. The extended vector of features for Vehicle $a$ becomes:

$$r_{f_a} = [r_{\delta_a} \quad \vdots \quad r_{p_{v_a}}]^T \tag{11}$$

Consequently, when considering Table 5, the extended rule base consequence vector $r_{f_a}$, or feature vector, comprises the individual partitioned components for peak deformation $\delta_a$ and pedestrian impact velocity $p_{v_a}$ and these are given, respectively, by:

$$r_{\delta_a} = [E_\delta(0.2681) \quad D_\delta(\delta_a) \quad C_\delta(\delta_a) \quad B_\delta(\delta_a) \quad A_\delta(0.8874)] \tag{12}$$

$$r_{p_{v_a}} = [E_{p_v}(6.7056) \quad D_{p_v}(p_{v_a}) \quad C_{p_v}(p_{v_a}) \quad B_{p_v}(p_{v_a}) \quad A_{p_v}(24.5872)] \tag{13}$$

The degrees of membership for the two component features to their pair-wise lower membership functions may be determined, respectively, from:

$$u_{\delta_{a_l}}(\delta_a) = \frac{\delta_{a_h} - \delta_a}{\delta_{a_h} - \delta_{a_l}} \tag{14}$$

$$u_{v_{a_l}}(p_{v_a}) = \frac{p_{v_{a_h}} - p_{v_a}}{p_{v_{a_h}} - p_{v_{a_l}}} \tag{15}$$

In each case the subscripts $l$ and $h$ correspond to the lower and higher adjacent membership functions, i.e., $E$ to $D$, $D$ to $C$, $C$ to $B$ and $B$ to $A$. The degrees of membership for the two component features to the corresponding pair-wise higher membership functions may then be determined, respectively, from:

$$u_{\delta_{a_h}}(\delta_a) = 1 - \mu_{\delta_{a_l}}(\delta_a) \tag{16}$$

$$u_{v_{a_h}}(p_{v_a}) = 1 - \mu_{v_{a_l}}(p_{v_a}) \tag{17}$$

The results of the injury severity from Equations (12) to (17), adopting generalised notation for the degrees of membership (i.e., $\mu_{\delta_l}, \mu_{\delta_h}, \mu_{p_{v_l}}, \mu_{p_{v_h}}$), are represented by the row-partitioned 2 x 2 matrix, denoted $x_{S_{AV}}$, given by:

$$x_{S_{AV}} = \begin{bmatrix} \mu_{\delta_l}(\delta_a) & \mu_{\delta_h}(\delta_a) \\ \mu_{p_{v_l}}(p_{v_a}) & \mu_{p_{v_h}}(p_{v_a}) \end{bmatrix} \tag{18}$$

where the injury severity will be represented by degrees of membership to two adjacent fuzzy sets at any point in time, i.e., $E$ to $D$, $D$ to $C$, $C$ to $B$ and $B$ to $A$. Note that the subscript $a$ which conveniently corresponds to Vehicle $a$ can be replaced by, e.g., $b, c$, to represent other AVs in a multiple vehicle scenario.

### 4.3 Common form Weighted Utility Cost Function for Collision Injury Severity Levels

This sub-section details the development of a novel common form of weighted nonlinear utility cost function for injury severity levels for the two potential collision outcomes detailed in Section 1.0, and illustrated in Figure 1. Note that these are of differing quantities and magnitudes; i.e., peak deformation of the vehicle in meters and pedestrian impact velocity being in meters/second. Prompted by this observation, the need for a common form of weighted utility cost function, for both collision outcomes, has been identified. The basis of the weighted utility cost function exploits the features in the fuzzy sets detailed in Section 4.1. Whilst each of the outcomes exists on their fuzzy set with equally spaced membership functions on a linear range, the weighted utility cost function is set-up in a nonlinear manner. The reason this is designed to be of a nonlinear form is an attempt to reduce utility costs by avoiding severe to fatal collisions. Also, by exploiting the fuzzy sets for each of the features, a common utility cost unit is readily derived whereby meaningful decisions can be made based on the numerical value of each cost function. This then results in a common utility cost being derived for the injury severity levels arising from peak deformation and pedestrian impact velocity.

The linear crisp values given in Equation (18) are used in this stage, i.e., the injury severity represented by degrees of membership to two adjacent fuzzy sets at any point in time. The linear crisp values for each feature are mapped on to a weighted cost function, whereby the crisp values from the lowest to highest map linearly on to, $n = 1 - 5$. In each case, these are used to populate the set of nonlinear factorial squared functions, denoted $N_{f_l}$ and $N_{f_h}$, with these given by:

$$N_{f_l} = \left(n_{f_l}!\right)^2 \tag{19}$$

$$N_{f_h} = \left(n_{f_h}!\right)^2 \tag{20}$$

where $f$ is used as the general subscript to represent the two feature indices of peak deformation and impact velocity, $n_{f_l}$ is based on the membership to the lower bounds and $n_{f_h}$ is based on the membership to the upper bounds. The purpose of using the factorial squared functions of Equations (19) and (20) is illustrated in Table 6, where the higher membership sets $n_{l/h}$ (i.e., $n$ being 4 and 5) give significantly higher values for both $N_{f_l}$ and $N_{f_h}$. As stated above, the intended aim of this common form of nonlinear cost function is to amplify the collision injury severity levels of severe collisions, with the aim then being to avoid such severe collisions, hence reduce the utility cost to society.

To define the utility cost function of lives at risk, the functions from Equations (19) and (20) are used together with the lower membership functions $u_{f_l}(f)$ and higher member functions $u_{f_h}(f)$. The general form for the weighted utility cost function representing lives at risk is given by:

$$\acute{\eta}(f) = \left(N_{f_h}(u_{f_h}(f)) + N_{f_l}(u_{f_l}(f))\right)N_{o/p} \tag{21}$$

where $N_{o/p}$ denotes the number of occupant(s)/pedestrian(s). Equation (21) now defines the common form the weighted utility cost functions of lives at risk for each feature, i.e., peak deformation and impact velocity for a given AV, denoted Vehicle $a$, which are given, respectively, by:

$$\acute{\eta}(\delta_a) = \left(N_{f_h}(u_{\delta_h}(\delta_a)) + N_{f_l}(u_{\delta_l}(\delta_a))\right)N_{o_a} \tag{22}$$

$$\acute{\eta}(p_{v_a}) = \left(N_{f_h}(u_{p_{v_h}}(p_{v_a})) + N_{f_l}(u_{p_{v_l}}(p_{v_a}))\right) N_{p_a} \tag{23}$$

As stated in Section 3, the point where deformation exceeds $0.5900m$, after which intrusion into the passenger compartment commences, and the value of pedestrian impact velocity exceeds $15.68m/s$ are required to be positioned on their respective universes of discourse to coincide with the peak of fuzzy set C, see Figure 9. It is crucial that these points are aligned in order for the numerical values of the common utility costs derived from Equations (22) and (23) to have a meaningful role to play within the ethical decision maker.

**Table 6:** *Factorial-Squared Function used to Distinguish the Collision Injury Severity Levels in a Common Weighted Nonlinear Manner*

| $n_{f_{l/h}}$ | $N_{f_{l/h}}$ |
|---|---|
| 1 | 1 |
| 2 | 4 |
| 3 | 36 |
| 4 | 576 |
| 5 | 14400 |

### 4.4 Ethical Decision Making Algorithms

The ethical decision-making algorithms used for collision target selection are now introduced. These are based on the philosophical approaches of Emanual Kant and Jeremy Bentham, i.e., the deontological and utilitarian approaches, respectively.

The deontological approach is based on the ethical views of Emanual Kant, see (Kant, 2017). Applying this approach to a collision scenario would ensure that the AV followed its natural path, hence without any intention to change paths to potentially save lives. Algorithm 1 describes the deontological approach that is to be evaluated in this research, where the AV effectively continues on its predetermined course/route without intervention.

**Algorithm 1:** *Deontological (Kant) Algorithm for Autonomous Vehicle Collision Target Selection*

i. Regardless of whether or not the AV is to be involved in an unavoidable collision, do not change the originally intended course/route

The utilitarian approach is based on the ethical views of Jeremey Bentham, see (Burns and Hart, 1998). This approach operates with the AV taking steering action to steer into the collision path that will save as many lives as possible or reduce the injury severity level of the collision outcome for the greater good. Steering action is taken even if this means that the AV occupant(s) might be sacrificed.

Algorithm 2 has been designed such that the AV will take the collision path with least injury severity. Such an approach aims to minimise the occupant/passenger injuries and potentially maximise the number of lives saved. Algorithm 2 will lead to intervention based on the initial information on collision injury severities captured in Section 3.0. The framework around this was then developed in Sections 4.1, 4.2 and 4.3, where weighted utility cost functions with a common structure for collision injury severity levels for AV peak deformation and occupant impact velocity have been developed, see Equations (22) and (23). Based on the values determined from Equations (22) or (23), the AV is commanded to steer onto the path with the lowest value for the utility cost of lives at risk, i.e., leading to the least overall utility cost to society.

**Algorithm 2:** *Utilitarian (Bentham) Algorithm for Autonomous Vehicle Collision Target Selection*

i. Obtain the collision injury severity, i.e., degree of membership to the fuzzy sets $A$, $B$, $C$, $D$ and $E$ for the two features, i.e., vehicle peak deformation $\delta_a$ and pedestrian impact velocity, $p_{v_a}$

ii. Determine the number of pedestrians, denoted $N_p$ and occupants, denoted $N_o$

iii. Assign an ID number, denoted $n$, where $n$ is based on the membership to the higher, denoted $n_h$ and lower bounds, denoted $n_l$ (i.e., $n \to 5, n \to 4, n \to 3, n \to 2$ and $n \to 1$) for the AV peak deformation and also the pedestrian impact velocity

iv. Determine the utility cost of lives at risk for the AV occupant(s) and pedestrian(s) using the following general form of the nonlinear utility cost function:

$$\acute{\eta}(f) = \left(N_{f_h}(u_{f_h}(f)) + N_{f_l}(u_{f_l}(f))\right)N_{o/p}$$

where $N_{f_l} = (n_{f_l}!)^2$ and $N_{f_h} = (n_{f_h}!)^2$

whereby the above expression for $\acute{\eta}(f)$ is used for the two collision features, $\delta_a$ and $p_{v_a}$ to give $\acute{\eta}(\delta_a)$ and $\acute{\eta}(p_{v_a})$

v. Steer into the path with the lowest value of the common utility cost, hence reduce injury severity levels and lives at risk, leading to the least utility cost to society

## 4.5 Illustrative Example

The illustrative example in this Section builds on that given in Section 2.5, where the collision output properties were as follows:

- Velocity: $17.32 m/s$
- Mass: $1407 kg$

Using the developed lumped parameter model in Section 3.2, the predetermined collision outcome results are given in Figure 10, where the peak deformation of the AV is determined to be $0.687 m$

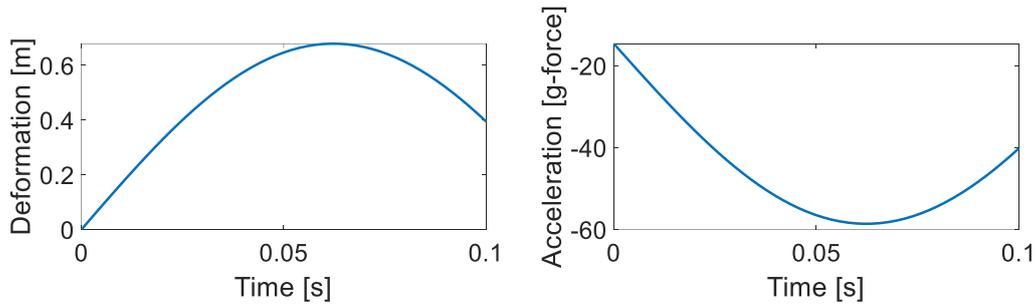

**Figure 10:** *Lumped Parameter Model for Deformation Versus Time (top-left hand sub-plot) and Acceleration versus Time (top-right hand sub-plot)*

Using the collision impact velocity for the pedestrian and the peak deformation for the AV, both of these are found to lie within their respective fuzzy sets $B$ to $C$ (the fuzzy set values given in Table 5, Section 4.1). Using Equations (16) to (17), the degrees of membership to the lower and upper fuzzy sets were determined, with the row-partitioned matrix, defined in Equation (18) now given by:

$$x_{S_{AV}} := \begin{bmatrix} \mu_{\delta_C}(\delta_a) & \mu_{\delta_B}(\delta_a) \\ \mu_{p_{v_C}}(p_{v_a}) & \mu_{p_{v_B}}(p_{v_a}) \end{bmatrix} := \begin{bmatrix} 0.2946 & 0.7054 \\ 0.6255 & 0.3745 \end{bmatrix} \qquad (24)$$

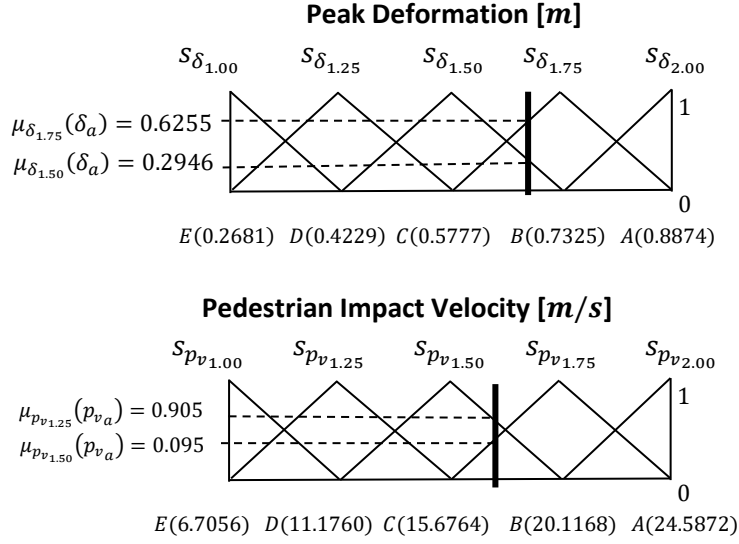

**Figure 11:** *Universe of Discourse for the Fuzzy Sets Corresponding to the Injury Severity for Peak Deformation $\delta_a$ and Pedestrian Impact Velocity $p_{v_a}$*

The degrees of membership to their respective fuzzy sets $B$ to $C$ are now used to determine the common utility cost via the factorial-squared function using values given in Table 6. In this case the values of $n$ presented initially in Equations (19) and (20) are given by 3 and 4 in this example, respectively. Using Equations (22) and (23), the values of utility cost of lives at risk are now determined, with the results given in Table 7. Based on the results given in Table 7, Algorithms 1 and 2, i.e., deontological and utilitarian ethical algorithms now are investigated. In the case of the deontological algorithm, the AV would remain on course and collide with the rigid barrier, with this being the lowest utility cost of lives at risk. In this particular case, the utilitarian algorithm would also lead to the AV taking no action and would collide into the rigid barrier.

**Table 7:** *Collision Outcome and the Common Utility Cost Results*

| Collision Outcome | Common Utility Cost |
| --- | --- |
| Rigid barrier (peak deformation) with two occupants | 390.2 |
| Two pedestrians (impact velocity) | 476.4 |

### 5.0 RESULTS FROM FURTHER SCENARIOS AND DISCUSSION

The ESCS will now be used to generate results based on the scenario given in Figure 1, Section 1.0. The following collision scenarios will be investigated:

- AV peak deformation (occupants zero) and AV impact velocity into pedestrian(s) (zero to four)
- AV peak deformation (occupants one) and AV impact velocity into pedestrian(s) (zero to four)
- AV peak deformation (occupants two) and AV impact velocity into pedestrians(s) (zero to four)

Note that an increase in the number of occupants from the laden mass of $1247kg$ (i.e., zero occupants), will result in the AV mass increasing. The increase in AV mass due to the occupants was detailed in Section 1.0, where an occupant mass of $80kg$ is used, i.e., one occupant will result in an AV laden mass of $1327kg$ and two occupants will result in an AV laden mass of $1407kg$. To recap from Section 1.0, AV initial velocities of $12m/s$, $16m/s$ and $20m/s$ are used in the simulation, and the number of pedestrians varies from 0 to 4. The initial simulation results relating to Section 2.0 (i.e., steering control system) are given in Table 8, with these detailed here for completeness. The data is then run through the algorithms developed in Section 4.0, with the results detailed in the following Sub-Sections.

The results from use of the developed ESCS coupled with the weighted nonlinear utility cost functions will be demonstrated. The results aim to show that when using the utilitarian approach, the AV will always select the collision path of least severity. This will correspond to the lowest value from either of the weighted utility cost functions given by Equations (22) and (23).

## 5.1 Collision Scenario 1: AV (Zero Occupants) and Pedestrian(s) (Zero to Four)

As the intention of the ESCS, when applying the utilitarian algorithm (see Algorithm 2, Section 4.4) is to steer onto the path of least severity, the results presented in this scenario are relatively straight forward. As there are zero occupants onboard the AV (i.e., zero risk), the AV will always steer into the rigid barrier. The results for this are given in Figure 12 for the three different initial AV velocities: $12m/s$ (Upper-Left Hand plot), $16m/s$ (Upper-Right Hand plot) and $20m/s$ (Lower-Left Hand plot). Table 8 contains the data used for the graphical outputs illustrated in Figure 12. As stated, as there are no occupants on-board the AV, hence, the common utility cost function value is zero. The AV will always self-sacrifice and collide into the rigid barrier. If the deontological algorithm was to be applied (see Algorithm, 1, Section 4.4), the collision outcome would be the same, i.e., the AV (containing zero occupants) and would remain on course and collide into the rigid barrier. However, consider the scenario whereby the AV is now in the top lane of Figure 1, Section 1.0. In such a scenario, applying the deontological approach could potentially result in injuries and/or fatalities.

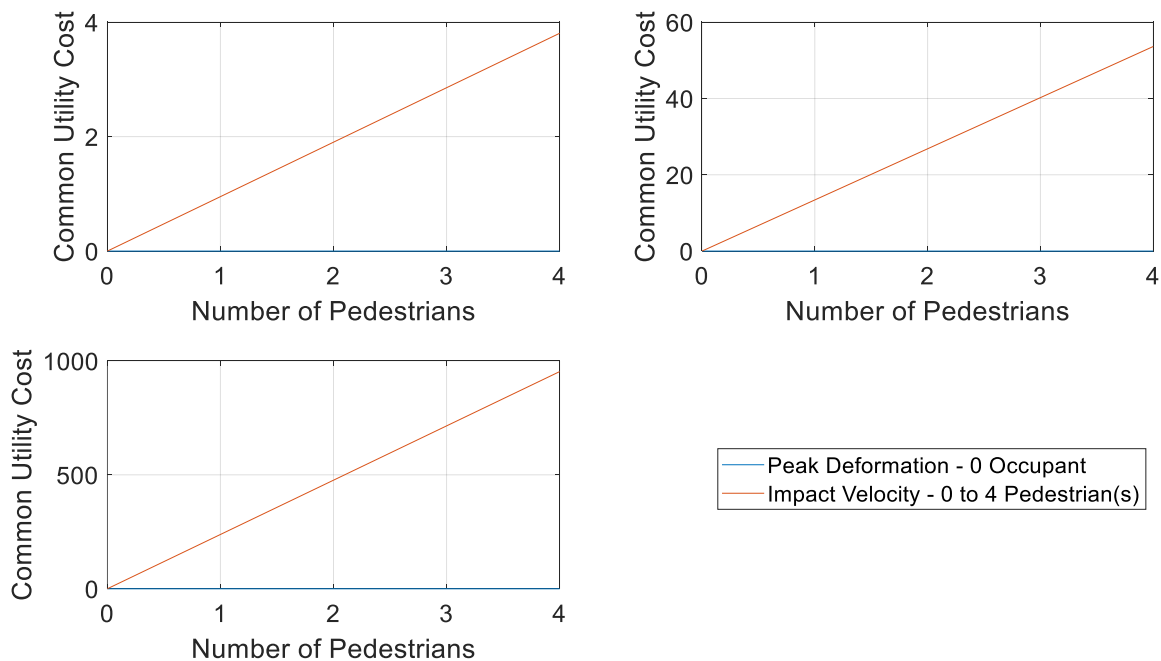

**Figure 12:** *Common Utility Cost for AV Peak Deformation (Zero Occupants) and AV Impact Velocity into Pedestrians (Zero to Four Pedestrian(s) for Three Velocities; 12m/s (Upper-Left Hand Plot), 16m/s (Upper-Right Hand plot) and 20m/s (Lower-Left Hand Plot)*

**Table 8:** *Common Utility Cost for Pedestrian(s) and Occupant for Varying Pedestrian Numbers, Zero Occupants and Impact Velocities*

| Pedestrian(s) | Occupant | Impact Velocity [m/s] | Common Utility Cost - Pedestrian(s) | Common Utility Cost - Occupant |
|---|---|---|---|---|
| 0 | 0 | 12 | 0 | 0 |
| 1 | 0 | 12 | 0.9514 | 0 |
| 2 | 0 | 12 | 1.9029 | 0 |
| 3 | 0 | 12 | 2.8543 | 0 |
| 4 | 0 | 12 | 3.8058 | 0 |
| 0 | 0 | 16 | 0 | 0 |
| 1 | 0 | 16 | 13.4058 | 0 |
| 2 | 0 | 16 | 26.8117 | 0 |
| 3 | 0 | 16 | 39.8117 | 0 |
| 4 | 0 | 16 | 53.6234 | 0 |
| 0 | 0 | 20 | 0 | 0 |
| 1 | 0 | 20 | 238.2231 | 0 |
| 2 | 0 | 20 | 476.4463 | 0 |
| 3 | 0 | 20 | 714.6694 | 0 |
| 4 | 0 | 20 | 952.8926 | 0 |

### 5.2 Collision Scenario 2: AV (One Occupant) and Pedestrian(s) (Zero to Four)

This will now investigate a scenario whereby the AV now contains one occupant. Figure 13 illustrates the results, where applying a utalitarian ethical approach (Algorithm 2 in Section 4.4) will involve the AV taking the path with the lowest value from the common utility cost functions, Equations (22) and (23). Table 9 contains the data used for the graphical outputs illustrated in Figure 13. Considering the initial AV velocity of $12m/s$ (Upper-Left-Hand plot) and $16m/s$ (Upper-Right-Hand plot), the lowest common utility cost unit is for peak deformation, i.e., the AV colliding into the rigid barrier. However, when the AV initial velocity is increased to $20m/s$, this result changes for the initial case whereby the AV makes a decision between colliding into one pedestrian or the rigid barrier. For such a scenario, the AV colliding into a pedestrian gives the lowest common utility cost. As the number of pedestrians is increased to two, the result changes, with the utilitarian ethical decision maker now selecting to collide into the rigid barrier. The AV will take the decision to steer into the rigid barrier for any scenario where there are two or more pedestrians, with an initial AV velocity of $20m/s$. If the deontological algorithm was to be applied (see Algorithm, 1, Section 4.4), for velocities of 12m/s and 16m/s the AV would be colliding into the rigid barrier (containing one occupant) in the case whereby there is zero pedestrians, i.e., in both cases the common utility cost is higher when colliding into the rigid barrier. However, for such velocities, the decision of colliding into the rigid barrier where there are one or more pedestrians is the path with the lowest injury severiy. Considering the 20m/s case, the common utility cost is higher in the case of the AV colliding in the barrier for zero and one pedestrian. However, for such velocities, colliding into the rigid barrier where there are two or more pedestrians is the path with the lowest injury severity.

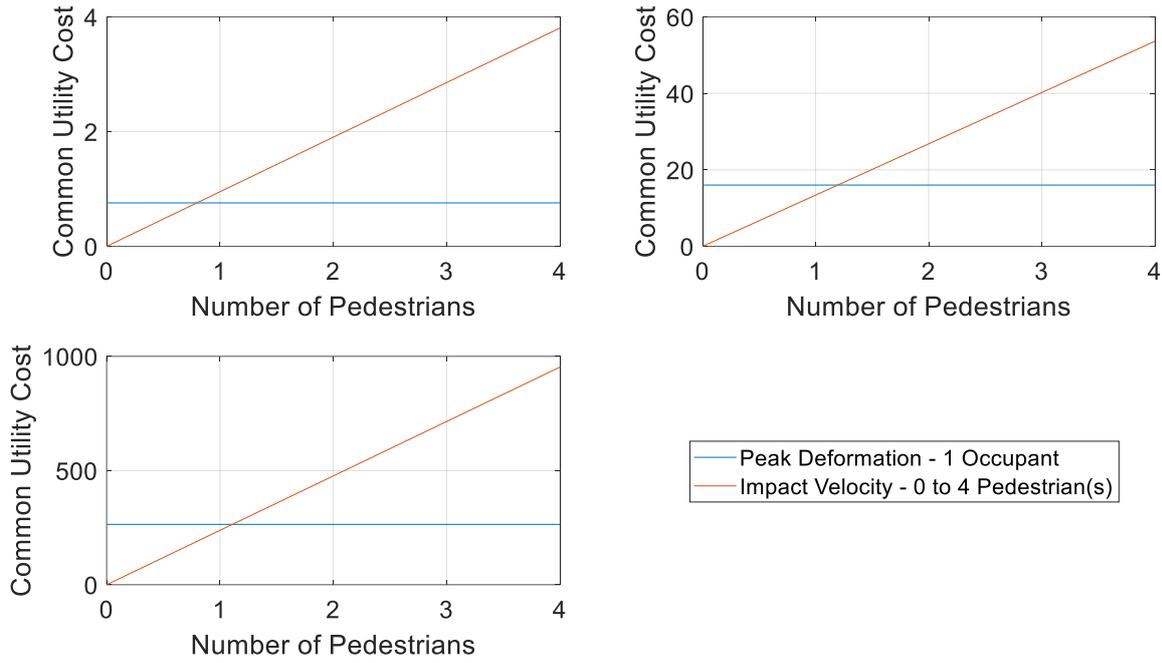

**Figure 13:** *Common Utility Cost for AV Peak Deformation (One Occupant) and AV Impact Velocity into Pedestrians (Zero to Four Pedestrian(s) for Three Velocities; 12m/s (Upper-Left Hand Plot), 16m/s (Upper-Right Hand plot) and 20m/s (Lower-Left Hand Plot)*

**Table 9:** *Common Utility Cost for Pedestrian(s) and Occupant for Varying Pedestrian Numbers, One Occupant and Impact Velocities*

| Pedestrian(s) | Occupant | Impact Velocity [m/s] | Common Utility Cost - Pedestrian(s) | Common Utility Cost - Occupant |
| --- | --- | --- | --- | --- |
| 0 | 1 | 12 | 0 | 0.7579 |
| 1 | 1 | 12 | 0.9514 | 0.7579 |
| 2 | 1 | 12 | 1.9029 | 0.7579 |
| 3 | 1 | 12 | 2.8543 | 0.7579 |
| 4 | 1 | 12 | 3.8058 | 0.7579 |
| 0 | 1 | 16 | 0 | 16.028 |
| 1 | 1 | 16 | 13.4058 | 16.028 |
| 2 | 1 | 16 | 26.8117 | 16.028 |
| 3 | 1 | 16 | 40.2175 | 16.028 |
| 4 | 1 | 16 | 53.6234 | 16.028 |
| 0 | 1 | 20 | 0 | 264.3827 |
| 1 | 1 | 20 | 238.2231 | 264.3827 |
| 2 | 1 | 20 | 476.4463 | 264.3827 |
| 3 | 1 | 20 | 714.6694 | 264.3827 |
| 4 | 1 | 20 | 952.8926 | 264.3827 |

### 5.3 Collision Scenario 3: AV (Two Occupants) and Pedestrian(s) (Zero to Four)

This will now investigate a scenario whereby the AV now contains two occupants. Figure 14 illustrates the results, where applying a utilitarian ethical approach will involve the AV taking the path with the lowest common utility cost. Table 10 contains the data used for the graphical outputs illustrated in Figure 14. Considering the initial AV velocity of 12m/s (Upper-Left-Hand plot), the AV will make the decision to steer into one pedestrian (when two occupants on on-board the AV). However, when there are two pedestrians, the AV will steer into the rigid barrier, with this decision repeating for two or more pedestrians. Now consideirng the 16$m/s$ (Upper-Right-Hand plot) and 20$m/s$ (Lower-Left-Hand plot) results in Figure 14, the same result occurs as with 12m/s, i.e., the AV will take action to steer into the case with one pedestrian, however, for more than one pedestrian the AV will select to steer into the rigid barrier. If the deontological algorithm was to be applied (see Algorithm, 1, Section 4.4) to this scenario, the outcome for all of the velocities is the same. The deontological approach would involve the AV colliding into the rigid barrier even when there is one pedestrian, even though the lower common utility cost would involve colliding into the one pedestrian.

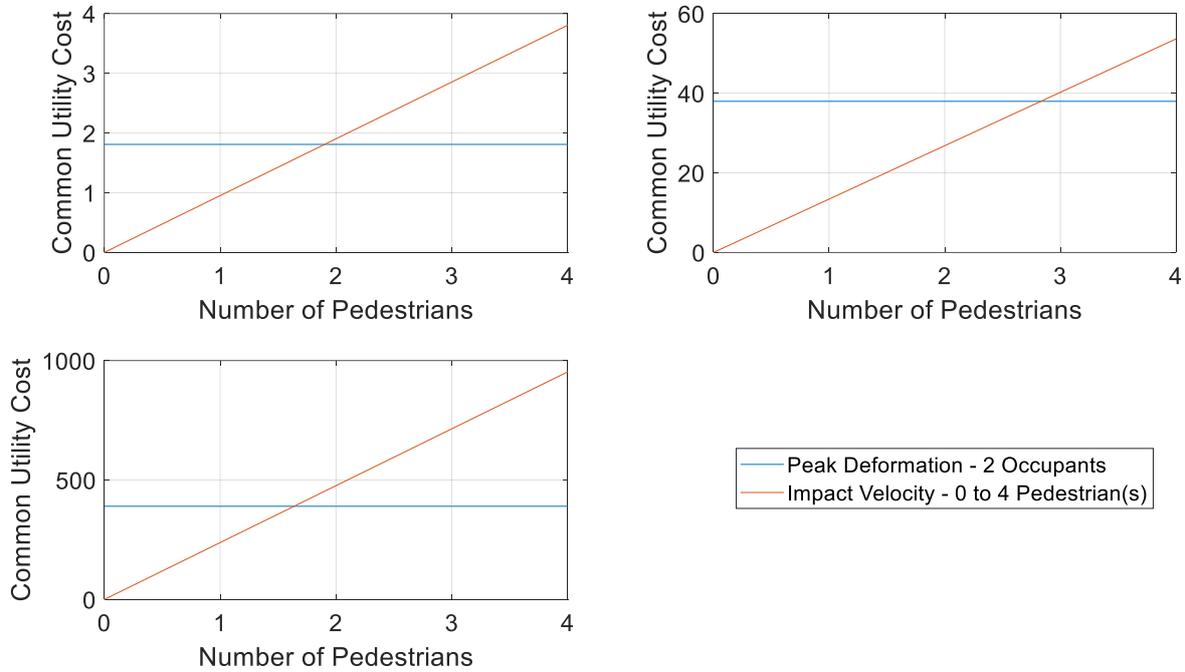

**Figure 14**: *Common Utility Cost for AV Peak Deformation (Two Occupants) and AV Impact Velocity into Pedestrians (Zero to Four Pedestrian(s) for Three Velocities; 12m/s (Upper-Left Hand Plot), 16m/s (Upper-Right Hand plot) and 20m/s (Lower-Left Hand Plot)*

**Table 10**: *Common Utility Cost for Pedestrian(s) and Occupant for Varying Pedestrian Numbers, Two Occupants and Impact Velocities*

| Pedestrian(s) | Occupant | Impact Velocity [m/s] | Common Utility Cost - Pedestrian(s) | Common Utility Cost - Occupants |
|---|---|---|---|---|
| 0 | 2 | 12 | 0 | 1.8092 |
| 1 | 2 | 12 | 0.9514 | 1.8092 |
| 2 | 2 | 12 | 1.9029 | 1.8092 |
| 3 | 2 | 12 | 2.8543 | 1.8092 |
| 4 | 2 | 12 | 3.8058 | 1.8092 |
| 0 | 2 | 16 | 0 | 37.9579 |
| 1 | 2 | 16 | 13.4058 | 37.9579 |
| 2 | 2 | 16 | 26.8117 | 37.9579 |
| 3 | 2 | 16 | 40.2175 | 37.9579 |
| 4 | 2 | 16 | 53.6234 | 37.9579 |
| 0 | 2 | 20 | 0 | 390.205 |
| 1 | 2 | 20 | 238.2231 | 390.205 |
| 2 | 2 | 20 | 476.4463 | 390.205 |
| 3 | 2 | 20 | 714.6694 | 390.205 |
| 4 | 2 | 20 | 952.8926 | 390.205 |

## 5.4 Quantifying the Collision Outcomes

Based on the Scenario presented in Figure 1, Section 1.0 and considering collision Scenarios 1, 2 and 3 in Sections 5.1 to 5.3, the common utility cost for the ethical decision-making algorithm outcomes for each scenario are summed, see Table 11. Hence, for each scenario a total of 15 common utility cost values are summed, i.e., for each of the 15 cases for the three scenarios, the common utility cost value is selected for either pedestrian(s) or occupant(s) (depending on the selected algorithm, i.e., utilitarian, or deontological). It is notable that for Scenarios 2 and 3, the common utility costs are lower values for the utilitarian approach, i.e., overall, the utilitarian ethical algorithm reduces the injury severity and/or fatalities, and consequently the utility cost to society. The deontological approach will always take the decision to collide into the rigid barrier, with this not always being the lowest value for the common utility cost (hence, the greater summed common utility cost values).

**Table 11:** *Summed Common Utility Cost Values from the Three Scenarios*

| Ethical Decision Maker Algorithm | Scenario 1: Common Utility Cost | Scenario 2: Common Utility Cost | Scenario 3: Common Utility Cost |
|---|---|---|---|
| Utilitarian approach | 0 | 1096 | 1531 |
| Deontological approach | 0 | 1406 | 2150 |

## 6.0 CONCLUSIONS

This paper has presented results from a study involving the development of an ethical steering control system (ESCS). The ESCS is configured such that algorithms can be tested to allow for the overall lowest utility cost to society to be selected, thereby leading to the least severe injuries and/or fatalities. The ESCS approach involved the development of models that would allow for path planning and ethical decision making to be predetermined based on knowledge prior to an imminent collision. The developed approach has been demonstrated through simulation to be successful, whereby knowledge of the collision outcomes were estimated based on models and the use of fuzzy logic. A common form of a weighted utility cost function has been developed using a novel nonlinear factorial squared concept in an attempt to avoid severe collisions to reduce injury severity levels, hence utility cost to society, and this has been shown to be successful. By exploiting a fuzzy sets approach, coupled with the developed common form of weighted cost function allows the injury severity levels from the peak deformation of the AV and the pedestrian impact velocity to be separately examined and compared numerically to obtain a common utility cost. It has been shown, in terms of the scenarios considered, that a common utility cost provides a meaningful input to the ethical decision maker in regard to path planning for an AV. In terms of the ethical decision maker, the efficacy of algorithms based the deontological and utilitarian approaches have also been investigated. It has been demonstrated that the utilitarian approach, as expected, results in reducing the utility cost to society by selecting the path leading to the least injury severity level for all the scenarios considered. These scenarios have included initial AV velocities, number of pedestrians and number of occupants. Based upon the findings presented here, the utilitarian approach, as an integral part of the ESCS, would appear to offer promising results in ethical decision making and path planning for future AV development.